\documentstyle[12pt,psfig]{article}
\begin{document}
\begin{center}
{\Large \bf Hidden evidence of non-exponential nuclear decay}
\vskip0.5cm
N. G. Kelkar${^1}$, M. Nowakowski${^1}$ and K. P. Khemchandani${^2}$\\
{$^1$\it Departamento de Fisica, Universidad de los Andes, 
Cra. 1E No.18A-10, Santafe de Bogota, Colombia} 
\\
{$^2$\it Nuclear Physics Division, Bhabha Atomic Research Centre, \\
Mumbai 400 085, India}
\end{center}

\begin{abstract}
The framework to describe natural phenomena at their basics being
quantum mechanics, there exist a large number of common global
phenomena occurring in different branches of natural sciences. 
One such global phenomenon is spontaneous quantum decay. 
However, its long time behaviour is experimentally poorly known. 
Here we show, that by combining two genuine quantum mechanical results, 
it is possible to infer on this large time behaviour, directly from 
data. Specifically, we find evidence for non-exponential behaviour 
of alpha decay of $^8$Be at large times from experiments.
\end{abstract}
%\pacs{03.65.Ta,03.65.Nk,23.60.+e}
\\

The decay law in quantum mechanics is necessarily non-exponential 
\cite{khalfin} and
deviates from a simple exponential form at small and large times (for
reviews see \cite{fonda,peres1,nakazato}). 
Recently, some experimental facts regarding the short time
behaviour have confirmed the expectations \cite{nature1}. Till today this is not the
case for the long time evolution.
Although this long tail of the survival probability of an unstable
quantum state is more often than not, not directly observable 
\cite{rutherford,butt,norman} 
(indeed, we are 
not aware of any such successful experiment), the information about 
resonance time evolution, in general, and the long time behaviour, 
in specific,  
should as a matter of
principle be encoded in the resonant scattering data $A + a \to 
{\rm resonance}
\to B + b$. Since energy is a dual variable to time, we can imagine that
the survival amplitude
\begin{equation} \label{amplitude}
{\cal A}_{\Psi}(t)=\langle \Psi \vert e^{-iHt}\vert \Psi\rangle = \langle
\Psi \vert \Psi (t) \rangle
\end{equation}
(related to the survival probability, 
$P_{\Psi}(t)={\vert \cal A}_{\Psi}(t)\vert^2 / \vert 
{\cal A}_{\Psi}(0)\vert^2$) 
can be written as a Fourier transform of an energy dependent quantity
which can be constructed on the premises of the scattering matrix, $S$. 
If this is so and the
$S$-matrix itself for a resonant reaction is extractable from the experiment, 
the information encoded in the scattering allows us to infer on the time
evolution of the resonance produced as an intermediate state 
in the process. To see this point clearly, let us first consider the 
method for the recasting 
of the survival amplitude as a Fourier transform of a spectral function. 
Ever since it was derived by Fock and Krylov \cite{fockrylov} 
it has been the basis for most of the investigations 
on quantum unstable systems.

Consider a resonance $R^*$ formed as an intermediate unstable state in
a scattering process such as $A + a \rightarrow R^* \rightarrow A + a$. 
Since the unstable state, $\vert \Psi \rangle$, cannot be an
eigenstate of the (hermitian) Hamiltonian, we expand it (assuming a 
continuous spectrum) in terms of the 
energy eigenstates of the decay products $A$ and $a$ as follows:
\begin{equation}\label{psi}
\vert \Psi \rangle =  \int dE \,\,a(E)\, \vert E \rangle
\end{equation}
where $\vert E \rangle$ is the eigenstate and $E$ the total energy of the 
system $A + a$.
Substituting now for $\vert \Psi \rangle$ in (\ref{amplitude}), we get,  
\begin{eqnarray}
A_{\Psi}(t) &=& \int dE'\, dE \,\,a^*(E')\, a(E)\, \langle E' 
\vert e^{(-iHt)} \vert E \rangle \\ \nonumber
 &=& \int dE' \,dE \,\,a^*(E') \,a(E)\, e^{(-iEt)} \,\delta(E - E')
\end{eqnarray}
The proper normalization of $\Psi$ tells us that $|a(E)|^2$ should
have dimension of ($1/E$) and hence can be associated with density of
states. In the Fock-Krylov method, 
${\cal A}_{\Psi}(t)$ then reads
\begin{equation} \label{fock}
{\cal A}_{\Psi}(t)=\int_{E_{\rm th.}}^{\infty} dE \,\,
\rho_{{ }_\Psi}(E) \,e^{-iEt}
\end{equation}
where $E_{\rm th.}$ is the minimum sum of the masses of the decay 
products \cite{note}. 
Since we have expanded 
the unstable state $\vert \psi \rangle$ in terms of the eigentstates, 
the spectral function
$\rho_{{ }_\Psi}(E)$, is a probability density
to find the eigenstates $\vert E \rangle$ in $\vert \Psi \rangle$, or, in other
words, it is the continuum probability density of states
in a resonance. Hence we can write, 
\begin{equation} \label{probdensity}
\rho_{{ }_\Psi}(E) =
{ d {\rm Prob}_{\Psi}(E) \over dE} =
\vert \langle E\vert \Psi \rangle\vert^2\,.
\end{equation}

If we succeed, in a second step, to extract $\rho_{{ }_\Psi}(E)$ from
experimental data, we can process this data by equation 
(\ref{fock}) to conclude, 
now from experiment, on the long tail of the survival amplitude of the 
resonance under consideration. This would then be the closest we can ever come
to pinning down the large times in a decay directly from experiment albeit
by an indirect method. 

Theoretically, many different forms of $\rho_{{ }_\Psi}(E)$ are 
available and many
have been used in a rather ad-hoc fashion. The then sometimes specious
results, especially with regard to the large times, are not always compatible
with each other. Secondly, mostly these are purely theoretical constructs
for which it is hard to find a bona fide connection with experiments. 
In all this, one 
important result originating from statistical mechanics, seems to have been 
overlooked, at least in connection with unstable states. In  
calculating the second virial coefficients $B$ and $C$ for the equation of states in a gas, $pV=RT[1 +B/V + C/V^2 +...]$, Beth and Uhlenbeck 
\cite{beth} (the derivation of their result is reproduced 
in \cite{huang}, see also \cite{dashen1,dashen2}) 
found that the
difference between the density of states with interaction, $n_l$, and without, 
$ n_l^{(0)}$, is given by the derivative of the scattering phase shift 
$\delta_l$ as, 
\begin{equation} \label{densitystates}
n_l(k_{\rm CM}) - n_l^{(0)}(k_{\rm CM})={2l+1 \over \pi}\,{d \delta_l (E_{\rm CM})
\over  d E_{\rm CM}}
\end{equation}
where $l$ is the angular momentum of the $l^{th}$ partial wave and 
$k_{\rm CM}$ and $E_{\rm CM}$ are the momentum and energy in the 
centre-of-mass system of the scattering particles, 
respectively. If a resonance is formed during the scattering process,
$E_{\rm CM}$ becomes the energy of the resonance in its rest frame. 
Certainly, the density of states and the
probability density in (\ref{probdensity}) are connected.
Note that in the absence of interaction, since no resonance can be produced, 
one would expect (\ref{probdensity}) to
be zero. Switching off the interaction (by, say, letting the
coupling constants go to zero), $n_l$ will not become zero, but tend to 
$n_l^{(0)}$ from above. 
Therefore, as long as $n_l-n_l^{(0)} \ge 0$ (this is always the case for an 
isolated resonance), we can write for the continuum probability density of
states of the decay products in a resonance,  
\begin{equation} \label{identification} 
{ d {\rm Prob}_{\Psi_l}(E_{\rm CM}) \over dE_{\rm CM}} ={\rm const.}\,
{d \delta_l (E_{\rm CM}) \over  d E_{\rm CM}}
\end{equation}
which is the sought after connection between data, here in the form of 
$\delta_l$, and the survival amplitude in (\ref{fock}).
This method is a general (i.e. without any further restrictions with the 
exception
of our belief in quantum mechanics)  
feasible tool for studying the time evolution of resonances 
from data, only if there are no overlapping resonances, i.e., if we handle one
isolated resonance. In reality, this is difficult to realize and in most cases
overlapping resonances will ``distort'' $d\delta_l/dE_{\rm CM}$, which
can then have several maxima and minima, even negative (see Fig.1). 
These negative regions
are bound to appear between resonances as noted by Wigner long ago in a
different context \cite{wigner}. 
The realistic situation of several overlapping resonances implies that
the identification (\ref{identification}) is operative starting
from threshold and extending over one resonance region, but often not beyond,
as far as real experiments are concerned.
However, one very useful feature remains when we 
restrict ourselves to large times. Large times correspond to small energies,
which implies that in order to experimentally extract information on this
region, all we need to know is $\delta_l$ at threshold and in 
the vicinity of the resonance. 
The exact form of how $\rho_{{ }_\Psi}(E)$ falls off at large $E$, well
beyond the resonance region, is not important 
to conclude on the behaviour of ${\cal A}_{\Psi}(t)$ as $t \to \infty$.  
This is the reason why we 
restrict ourselves to large time behaviour in applying the method of Fock 
and Krylov with the inclusion of the result of Beth and Uhlenbeck.

To demonstrate the connection between the spectral density and 
phase shift derivative as in (\ref{identification}), we quote a simple 
example below. An amplitude which describes the resonant scattering 
process around the pole is often taken as,
\begin{equation}\label{tmat}
T = {\Gamma/2 \over E_R \,- \,E \,- \,i\Gamma/2}\,,
\end{equation}
from which one easily gets \cite{npaours}, 
\begin{equation}\label{9}
{d\delta \over dE} = {\Gamma/2 \over (E_R - E)^2 + {1 \over 4} \Gamma^2}\, ,
\end{equation}
i.e., the Lorentzian (Breit-Wigner) form. The right hand 
side of (\ref{9}), up to a constant, is commonly taken as the spectral 
function in (\ref{fock}) to display the fact that a one-pole 
approximation as in (\ref{tmat}) leads to the exponential decay law
\cite{sakurai,raczka,marek}. Note that we obtained this spectral function
as a derivative of the scattering phase shift. 
If a resonance, $R^*$, is produced an as intermediate state in the scattering 
of two particles, $A$ and $a$ as, $A + a \rightarrow R^* \rightarrow A + a$, 
then the energy derivative of the scattering phase shift $\delta$ for this
reaction, gives the spectral function for the unstable state or resonance
$R^*$. If we compute (\ref{9}) in the rest frame of the resonance, then  
the energy $E$ in (\ref{9}) is the energy available in the centre of mass
of the scattering particles $A$ and $a$. 
In the present work, we use the energy derivative of experimental 
scattering phase shifts to obtain the spectral
function corresponding to the intermediate unstable state in scattering.

Suppose now, that we have a fit to the data of $\delta_l$ with a reasonable
scan of the threshold/resonance region which will determine the large time
behaviour and which to know is therefore a conditio sine qua non. 
We said already that in calculating the survival probability for large times,
the large energy behaviour of every single resonance, in the situation
where several resonances are overlapping is not of importance. We now make
this statement more precise gaining as a byproduct more insight
into the late time domain. Since the phase shift $\delta_l$ has a
threshold behaviour, so will $\rho_{{ }_\Psi}$. We can parametrize it
without loss of generality in the form $\rho_{{ }_\Psi} \propto (E_{\rm CM}
-E_{\rm th.})^{\gamma}$ to account for the threshold. Hence we have
\begin{equation} \label{general1}
\rho_{{ }_{\Psi_l}}(E_{\rm CM})
={\cal G}(E_{\rm CM})(E_{\rm CM}-E_{\rm th.})^{\gamma (l)}\,.
\end{equation}
We impose the following condition on ${\cal G}(E_{\rm CM})$: (i) 
${\cal G}(E_{\rm th.}) \neq 0$ since we have factorized the 
threshold already, (ii) ${\cal G}(E_{\rm CM}) \to
0$ sufficiently fast as $E_{\rm CM} \to \infty$ (in case of several
overlapping resonances this is a theoretical assumption which,
however, is inherent in the Fock-Krylov method) and (iii) mathematically, 
we allow the function
to have poles $z_{0i}$ in the complex plane, i.e. $1/{\cal G}(z_{0i})=0$ 
such that $\Im m(z_{0i}) < 0$ and $\Re e(z_{0i}) > 0$; since the derivative
of the phase shift carries the information about the poles 
\cite{brans,peres2}
we would expect on physics grounds a single pole at $E_R -i\Gamma/2$
signifying the resonance parameters. 
This generalizes the simple Breit-Wigner form in (\ref{9}) which has the
pole but no threshold behaviour. 
These general properties allow 
us to compute
the survival probability by going to the complex plane 
(though this method of finding the survival probability is standard and
known \cite{fonda,nakazato}, we describe it here as it is not exactly
equivalent to that in \cite{fonda,nakazato,acta}).  
We choose the closed path
$C_{\rm R}= C_{\Im} + C_{\Re} + C_{\rm R}^{1/4}$, starting from
zero (after change of variables $y=E_{CM}-E_{th}$) along the 
real axis ($C_{\Re}$) attaching to it a quarter of a circle with
radius ${\rm R}$ ($C_{\rm R}^{1/4}$) in the clockwise direction and completing 
the path by going upward the
imaginary axis up to zero ($C_{\Im}$). In the integral 
we let ${\rm R}$ go to infinity noting
that along $C_{{\rm R} \to \infty}^{1/4}$, the property (ii) 
and the fact that $e^{-izt} \propto e^{\Im m z}$ ($\Im m z \le 0$), 
ensures that
the integral is zero. We subtract the contribution along the
imaginary axis. This gives
\begin{equation} \label{sum}
{\cal A}_{\Psi_l}(t)={\cal A}_{\Psi_l}^{\rm E}(t) 
+ {\cal A}_{\Psi_l}^{\rm P}(t)
\end{equation} 
with
\begin{equation} \label{expA}
{\cal A}_{\Psi_l}^{\rm E}(t)=e^{-iE_{\rm th.}t}\lim_{{\rm R}\to \infty}
\oint_{C_{\rm R}}dz e^{-izt} z^{\gamma} {\cal G}(z+ E_{\rm th.})=C_1\, 
e^{-iE_Rt}e^{-\Gamma/2\,t}
\end{equation}
by Cauchy's theorem and
\begin{eqnarray} \label{corrA}
{\cal A}_{\Psi_l}^{\rm P}(t)&=&C_2 \,e^{-iE_{th.}t} 
\int_0^{\infty}dx e^{-xt} x^{\gamma}
{\cal G}(-ix +E_{\rm th.}) \nonumber \\
&\simeq& C_2\, e^{-iE_{th.}t} {\cal G}(E_{\rm th.})
\Gamma (\gamma +1)\, {1 \over t^{\gamma +1}}
\end{eqnarray}
for the integral along $C_{\Im}$, 
where the approximation is valid for large times $t$. 
In the above, $\Gamma (x)$ is the Euler's gamma function and
$C_1$ is a constant easily
calculable in terms of the parameters $E_R$, $\Gamma$ and $E_{\rm th.}$ and 
$C_2$ is $(-i)^{\gamma + 1}$. 
Equations (\ref{expA}) and
(\ref{corrA}) show the general features which we would expect: an
exponential decay law followed by inverse power law corrections. The latter is
independent of the details of ${\cal G}(E_{\rm CM})$ 
as claimed, displaying also 
nicely the dual nature of time and energy. Formulae (\ref{expA}) and
(\ref{corrA}) substantiate our previous claims about the method to extract
information on large time behaviour of resonances even if the data
have several overlapping structures. To be mathematically precise, we 
note that the way we derived (\ref{corrA}) is valid only for integer 
values of the exponent $\gamma$. We refer the reader to \cite{nakazato}
for the general case, where, however, a similar result with the 
survival amplitude being proportional to $1/t^{\gamma + 1}$ is obtained
for large times.   

To make a direct contact with experiment we
need experimental data on $\delta_l$ which, parametrized, can be used to
extract the survival amplitude. However, the data have to start
very close to the threshold. This is not always easy to find in literature, 
as threshold regions are not the most interesting regions to focus on in an 
experiment. 

We opted for an experiment with many data points at threshold and relatively
small error 
bars. It is the $\alpha$-$\alpha$ D-wave resonant scattering in nuclear 
physics [17-21]
%\cite{hydenburg}, \cite{tombrello}, \cite{nilson}, \cite{chien}, \cite{bacher}
%\begin{equation} \label{alpha}
$$\alpha + \alpha \to { }^8{\rm Be}(2^+) \to \alpha + \alpha \,. $$
%\end{equation}
In Fig. 1 we display the phase shift and its derivative over a wide
region, using a simple polynomial fit to the phase shift. We find 
the established $^8$Be levels, shown in the figure.
Motivated by a Lorentzian form \cite{joachain} with an energy dependent 
width 
$\Gamma(E_{\rm CM})$, we parametrized the data 
in the region of the first $2^+$ resonance, in the following form:
\begin{equation} \label{para}
\delta_l(E_{\rm CM})=\tan^{-1}\left[{\Gamma(E_{\rm CM}) \over E_0 -E_{\rm CM}}
\right]\,e^{-\beta E_{\rm CM}}
\end{equation}
with
\begin{equation} \label{gamma}
\Gamma(E_{\rm CM})= \Gamma_0\left({E^2_{\rm CM} -E^2_{\rm th.} \over
E^2_0 -   E^2_{\rm th.}}\right)^{\kappa/2}
\end{equation}
\begin{figure}[h]
\centerline{\vbox{
\psfig{file=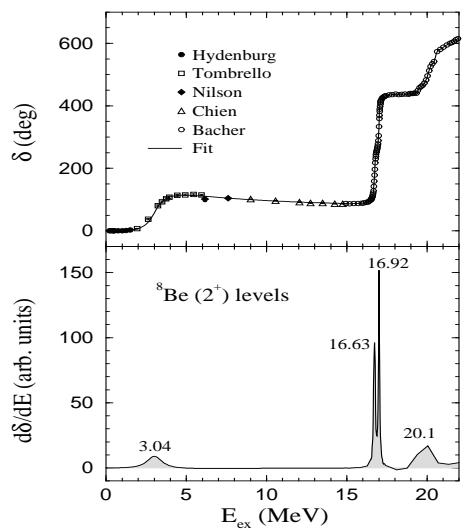,height=7cm,width=6cm}}}
\caption{
D-wave phase shifts (upper half) in $\alpha$-$\alpha$ elastic scattering
from refs [17-21], polynomial fit to these data
(solid line) and the derivative of phase shift (lower half) calculated from
the fit showing all established ${ }^8{\rm Be} (2^+)$ levels,
as a function of the
excitation energy $E_{ex} = E_{CM}- E_{{ }^8{\rm Be}{\rm (ground state)}}$
and plotted here in arbitrary units (arb. units).
The negative
region in derivative of phase shift (lower half) between $5-15$ MeV, due
to the slowly falling phase shift is not obvious in the plot due to
the scale of the vertical axis.}
\end{figure}
\begin{figure}[h]
\centerline{\vbox{
\psfig{file=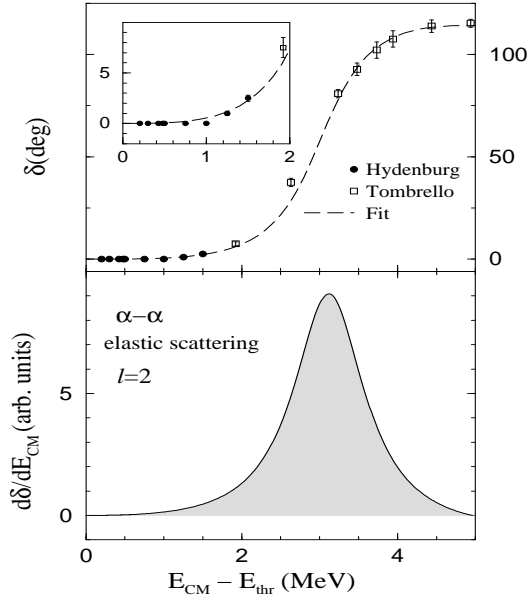,height=8cm,width=7cm}}}
\caption{
D-wave phase shift (upper half) and its derivative (lower half) in
$\alpha$-$\alpha$ elastic scattering as a function of $E_{CM} -
E_{\rm threshold}$,
in the region of the first $2^+$ level of ${ }^8{\rm Be}$.
The dashed line shows the fit mentioned in the text. The inset
displays the accuracy of the fit near the threshold energy region which
is crucial for the large time behaviour of the decay law.}
\end{figure}
\noindent
which is valid for the elastic case. The derivative of this parametrized
phase shift satisfies our requirements. In particular, $\kappa=2\gamma +2$
and the pole is highlighted 
by a peak structure (see Fig. 2) which is expected \cite{brans}. 
To be unbiased, we fitted $\kappa$, $\beta$, $\Gamma_0$
and $E_0$ simultaneously taking the error bars into consideration. 
Our best fit gives
$\kappa =6.36$, $\beta=0.00359\,\, {\rm GeV}^{-1}$, 
$\Gamma_0 =0.0009\,\, {\rm GeV}$ and $E_0=7.45838\,\, {\rm GeV}$ and is shown
also in Fig. 2. In the fitting procedure, special attention was paid to the
mandatory threshold.  
We can take the parametrization, the derivative of the same and 
perform numerically the integration to obtain the survival amplitude. 
The numerical result for the survival probability is depicted in 
Fig. 3 and is our main result.
However, since we see our
theoretical conditions (i)-(iii) on the spectral function confirmed, it
suffices to use the fit together with the mathematics developed in 
formulae (\ref{expA})
and (\ref{corrA}). 
We then conclude  
that at large times, the survival probability of the unstable 
${ }^8 {\rm Be}(2^+)$ state at $3.04$ MeV excitation energy, behaves as, 
\begin{equation} \label{concl}
P_{{ }_{ }^8{\rm Be}}(t) \sim { 1 \over t^{6.36}}\,.
\end{equation}
Theoretically we would expect that near threshold \cite{joachain,taylor},
%\begin{equation} \label{theory1}
$\tan \delta_l(E_{\rm CM}) \sim k_{\rm CM}^{2l+1}$, 
%\end{equation}
which implies, 
%\begin{equation} \label{theory2}
$d \delta_l (E_{\rm CM})/d E_{\rm CM} \sim (E_{\rm CM}-E_{\rm th.})^{l-1/2}$
%\end{equation}
also near threshold. 
\begin{figure}[h]
\centerline{\vbox{
\psfig{file=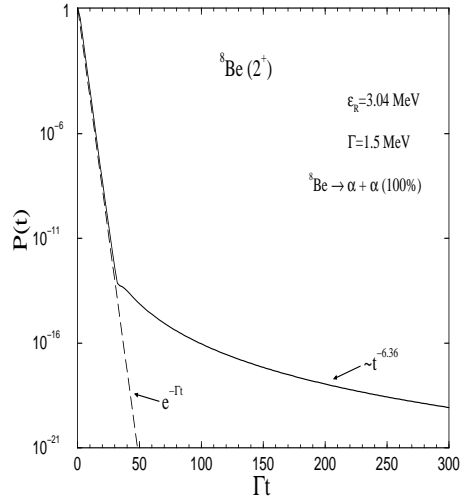,height=7cm,width=6cm}}}
\caption{
Survival probability $P(t)$ of the decay
${ }^8{\rm Be} (2^+) \rightarrow \alpha + \alpha$, as a function of number
of lifetimes after decay. $P(t) = {\vert \cal A}_{\Psi_l}(t)\vert^2 / \vert
{\cal A}_{\Psi_l}(0)\vert^2$ is evaluated numerically
using $d\delta_l/dE_{\rm CM}$ of Fig. 2
(which has been obtained from the fit to phase shift data) as the spectral
density $\rho_{{ }_\Psi}(E)$ in (4).
The dashed line ($e^{-\Gamma t}$) shows that
the decay law for the ${ }^8{\rm Be}(2^+)$ state (solid line) is exponential
up to about 30 lifetimes after which it proceeds as $t^{-6.36}$.
$\epsilon_{R}$ and $\Gamma$ are the resonance mass and width respectively.}
\end{figure}
This amounts to saying that $\kappa$ is expected to be $2l+1$.
For the $^8{\rm Be} (2^+)$ resonance, one then gets an inverse power law
$t^{-5}$ for the survival probability. The data on the phase shift do not
seem to follow the standard threshold behaviour and hence we get 
(\ref{concl}). The discrepancy, however,  
does not look serious. Indeed, re-calculated the ``$l$''-value
of the fitted $\kappa$ is 2.68. Interestingly, the exponent $6.36$ is close
to the theoretical prediction of $7$ for the case of $l=2$ made in 
\cite{fonda}. The deviation from the exponential decay law starts 
around 30 lifetimes after the onset of the decay. By this time, 
one could say that the sample with which one started has depleted
by about 13 orders of magnitude ($\sim e^{-30}$), making a direct
measurement of such a phenomenon not feasible. The above is the case
of a strong decay with short lifetime. In the case of weak decays,
the onset of the non-exponential law at large times is expected to
be much later \cite{fonda,peres1} making the direct measurement even less
feasible.

In summary, we combined the Fock-Krylov method to calculate 
the survival amplitude of an 
unstable state in terms of a Fourier transform of a spectral function with a 
result in statistical mechanics of Beth-Uhlenbeck which identifies the 
continuum density of states with the energy derivative of the two body 
scattering phase shift (\ref{densitystates}) being proportional to the 
continuum probability density of states (\ref{probdensity}).
Hence, using experimentally determined phase shifts, the method allowed us to
compute the non-exponential long time behaviour of unstable quantum systems
directly from data. An explicit example was given and the inverse power law 
behaviour at large times determined from data. Asked as to why the 
merger of the two
quantum mechanical results, which when combined give insight into the 
quantum decay, 
has been overlooked so far, we can only speculate
by answering that even the old results by Eisenbud \cite{eisen},
Wigner \cite{wigner} and Smith \cite{smith} concerning
the phase shift derivative have been neglected for a long time 
and came back into vogue only recently [30-34].
%\cite{hic,neels,neels2,neels3,chaos}.

We close by quoting from \cite{nature2}; ``Thus it seems unlikely that 
nuclear decays 
will show deviations from the exponential decay law which they made famous.''
We have shown that it is possible, as the information on the time evolutions
is encoded in the scattering data.


\begin{thebibliography}{99}
\bibitem{note}
In principle, $E_{th}$ is the lower bound of the total Hamiltonian. 
If one makes an expansion in terms of energy eigenstates of the 
decay products, then $E_{th}$ is the minimum sum of the masses of the
decay products.
\end{thebibliography}
\end{document}